\documentclass[12pt,preprint]{aastex}

\newcommand{\msun}{\mbox{M$_{\odot}$}}

\newcommand{\zsun}{\mbox{Z$_{\odot}$}}
\slugcomment{}

\shorttitle{Metallicity of HE 0437--5439}
\shortauthors{Bonanos et al.}

\begin{document}

\title{Low Metallicity Indicates that the Hypervelocity Star \\ HE
0437--5439 was Ejected from the LMC\altaffilmark{1}}

\author{Alceste Z. Bonanos\altaffilmark{2}, Mercedes
L\'opez-Morales\altaffilmark{3}}

\affil{Carnegie Institution of Washington, Department of
Terrestrial Magnetism, \\ 5241 Broad Branch Rd. NW, Washington D.C., 20015,
USA \\ \tt e-mail: bonanos@dtm.ciw.edu, mercedes@dtm.ciw.edu}

\author{Ian Hunter \& Robert S. I. Ryans}
\affil{Department of Physics and Astronomy, The Queen's
  University of Belfast, \\ BT7 1NN, Northern Ireland, UK \\ \tt e-mail:
  I.Hunter@qub.ac.uk, R.Ryans@qub.ac.uk}

\altaffiltext{1}{Based on observations made with the European Southern
Observatory telescopes obtained from the ESO/ST-ECF Science Archive
Facility and with the 1 meter Henrietta Swope telescope located at Las
Campanas Observatory, Chile.}  
\altaffiltext{2}{Vera Rubin Fellow.}
\altaffiltext{3}{Hubble Fellow.}

\begin{abstract}

We measure the metallicity of the unusual hypervelocity star HE
0437--5439 from high resolution spectroscopy to be half-solar. We
determine a spectral type of B2 IV-III for the star and derive an
effective temperature T$_{eff}=$ 21,500 $\pm$ 1,000~K and a surface
gravity $\log(g)$ = 3.7 $\pm$ 0.2 (cgs).  We also present $BV$ time
series photometry and find the star to be non-variable at the 0.02 mag
level. We refine the magnitude of the hypervelocity star to
$V=16.36\pm0.04$ mag, with a color $B-V=-0.23\pm0.03$ mag, confirming
its early-type nature. Our metallicity result establishes the origin of
HE 0437--5439 in the Large Magellanic Cloud and implies the existence of
a massive black hole somewhere in this galaxy.

\end{abstract}


\section{Introduction} \label{sec:intro}

\citet{Hills88} first proposed the existence of hypervelocity stars
(HVSs) as evidence for a supermassive black hole in the center of our
Galaxy. According to his calculations, an encounter of a close binary
with a supermassive black hole could disrupt the binary, capturing one
of the stars and ejecting the other as a HVS with a velocity up to 4000
km s$^{-1}$. Observational evidence for such objects was first presented
by \citet{Brown05}, who serendipitously discovered the first HVS, SDSS
J090745.0+024507, in a survey of blue horizontal branch
stars. \citet{Fuentes06} resolved the degeneracy between luminosity and
distance for this star by detecting variability and providing evidence
for the main-sequence nature of the first HVS.

To date ten HVSs have been reported \citep[][Table 1]{Brown07} and of
these, HE 0437--5439 stands out as the most enigmatic one. While the
origin of the other nine is consistent with being ejected from the
center of our Galaxy, HE 0437--5439 is suspected to have been ejected
from the Large Magellanic Cloud (LMC). \citet{Edelmann05} discovered
this HVS during a spectroscopic follow-up of their sample of subluminous
B-star candidates from the Hamburg/ESO survey
\citep[e.g.][]{Christlieb01}. HE 0437--5439, with an estimated apparent
magnitude of $V=16.2\pm 0.2$ mag, is the brightest HVS
known. \citet{Edelmann05} obtained two high resolution spectra, but with
a low signal to noise (S/N) ratio ($\sim$20), finding it to be a
main-sequence early B-type star consistent with solar metallicity. They
measured a heliocentric radial velocity of $+723\pm3$ km s$^{-1}$, a
rotational velocity of $54\pm4$ km s$^{-1}$ and derived a distance of
$d=61\pm12$ kpc from the Galatic center, which is inconsistent with the
spectral type of the star. While an early-type B-star has a
main-sequence lifetime of 25-35 Myr, it requires 100 Myr to travel that
distance at the measured radial velocity.

The authors therefore propose two possible explanations for this
paradox: either HE 0437--5439 is a Galactic blue straggler or it
originated from the LMC. They suggest measurements of the chemical
abundance and proper motion of the star to distinguish between the two
scenarios. \citet{Gualandris07} discarded the blue straggler scenario
claiming that a merger product would not live much longer than a
main-sequence star of the same mass. Assuming HE 0437--5439 was ejected
by a black hole in the LMC, they perform scattering simulations and
conclude that a black hole mass $\ge1000\;\msun$ is required to explain
the velocity of HE 0437--5439. 

Motivated by these intriguing scenarios, we set out to measure the
metallicity of HE 0437--5439 and obtained photometry to search for
variability. We describe our findings in the following sections of this
Letter.

\section{Photometry} \label{sec:phot}

\begin{figure}[hbt]
\epsscale{0.8}
\plotone{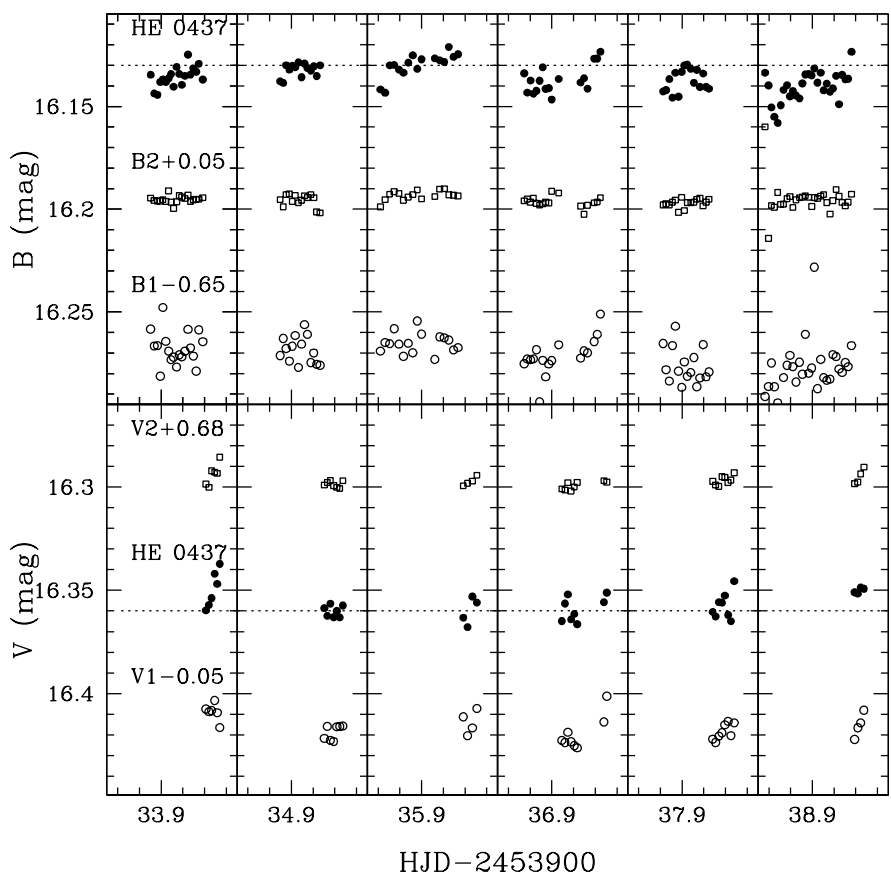}
\caption{$B$ and $V-$band light curves of HE 0437--5439 ({\em filled
circles}) and 2 other comparison stars of similar brightness, shifted
for display purposes. Panels correspond to each of the six consecutive
nights of observation; tickmarks on the x-axis correspond to 0.02 day
intervals. Dotted lines indicate the absolute photometry derived from
the second, photometric night: $V=16.36\pm0.04$ mag, $B-V=-0.23\pm0.03$
mag.}
\label{fig:lcbv}
\end{figure}

We imaged HE 0437--5439 in Johnson $B$ and $V$ bands with the Direct CCD
on the 1-m Swope telescope at Las Campanas Observatory, Chile. We
monitored the star over six consecutive nights (UT 2006 July 17-22), for
about 2 hours per night. The images were processed with standard
IRAF\footnote{IRAF is distributed by the National Optical Astronomy
Observatory, which are operated by the Association of Universities for
Research in Astronomy, Inc., under cooperative agreement with the NSF.}
routines, following the same procedure as described in
\citet{Bonanos07}. The light curves of HE 0437--5439 and several
reference stars of similar brightness in the field were extracted with
the ISIS image subtraction package \citep{Alard98,Alard00} in each
filter and converted to magnitudes, following \citet{Hartman04}. The $B$
and $V$ time series photometry for HE 0437--5439 and two of the
reference stars are shown in Figure \ref{fig:lcbv}. There is no evidence
for variability to a precision of 0.02 mag over the 6 nights. HE
0437--5439 is therefore not pulsating as SDSS J090745.0+024507
\citep{Fuentes06}, and the scenario of a blue straggler consisting of a
contact binary is also unlikely.

To calibrate the $B$ and $V$ magnitudes of HE 0437--5439, we performed
aperture photometry on images obtained on the photometric night of UT
2006 July 19 and applied transformation coefficients derived by
\citet{Bonanos07} for this night, measuring $V=16.36\pm0.04$ mag,
$B-V=-0.23\pm0.03$ mag. This improves the photometry measured by
\citet{Edelmann05} from the Hamburg/ESO plates ($V=16.2\pm0.2$ mag).
The $B-V$ color we measure is consistent with that of an early-type B
star. We note that the foreground extinction estimate from
\citet{Schlegel98} is $E(B-V)=0.008$ mag, therefore the true unreddened
$V_0$ and $(B-V)_0$ values are identical within the errors.

\section{Spectroscopy} \label{sec:spec}

We retrieved public, unpublished spectra of HE 0437--5439 from the ESO
archive. The data were obtained on UT 2006 January 12 with the
Ultraviolet and Visual Echelle Spectrograph (UVES) on the VLT UT2
8-meter telescope (Kueyen) at the ESO Paranal Observatory. The
observations consist of 8$\times1482\,s$ and 1$\times1810\,s$ spectra,
with a resolving power $R\sim34,000$, measured from the full width half
maximum of the comparison lamp lines. We bias subtracted and flatfielded
the spectra with the IRAF echelle package routines. We used the
algorithm of \citet{Pych04} to remove cosmic rays from each two
dimensional image. Next, we extracted the spectra, averaged them
(weighting by the exposure time), normalized and merged the orders. The
final spectrum ranges from 3746-4990\,\AA\, and has an average S/N ratio
of $\sim$100.

We classify HE 0437--5439 as a B2 IV-III star according to the criteria
defined by \citet{Walborn90}. The depths of the Si lines constrain the
spectral type to B1-B2. Furthermore, \ion{Si}{3} $\lambda$4552 $>$
\ion{Si}{4} $\lambda 4089$ and \ion{Si}{2} $\lambda \lambda$4128-30 $<$
\ion{Si}{3} $\lambda 4552$, while the \ion{O}{2} blends at $\lambda4640$
and $\lambda4650$ have near equal depths, in support of a B2 type
(\ion{C}{3} lines affect the relative depths in B1 types). The
luminosity class criteria are based on the relative strengths of the
\ion{He}{1} and \ion{Si}{3} lines: \ion{He}{1} $\lambda
4387/$\ion{Si}{3} $\lambda 4552\sim5$, corresponding to IV, and
\ion{He}{1} $\lambda\lambda$4144/4121 $\sim2$, corresponding to
IV-III. Our adopted spectral type, B2 IV-III, gives a slightly more
evolved star than what \citet{Edelmann05} found from their much lower
quality spectrum.

\section{Metallicity and Parameter Determination} \label{sec:metal}

\begin{figure}[hbt]
\epsscale{1.0}
\plotone{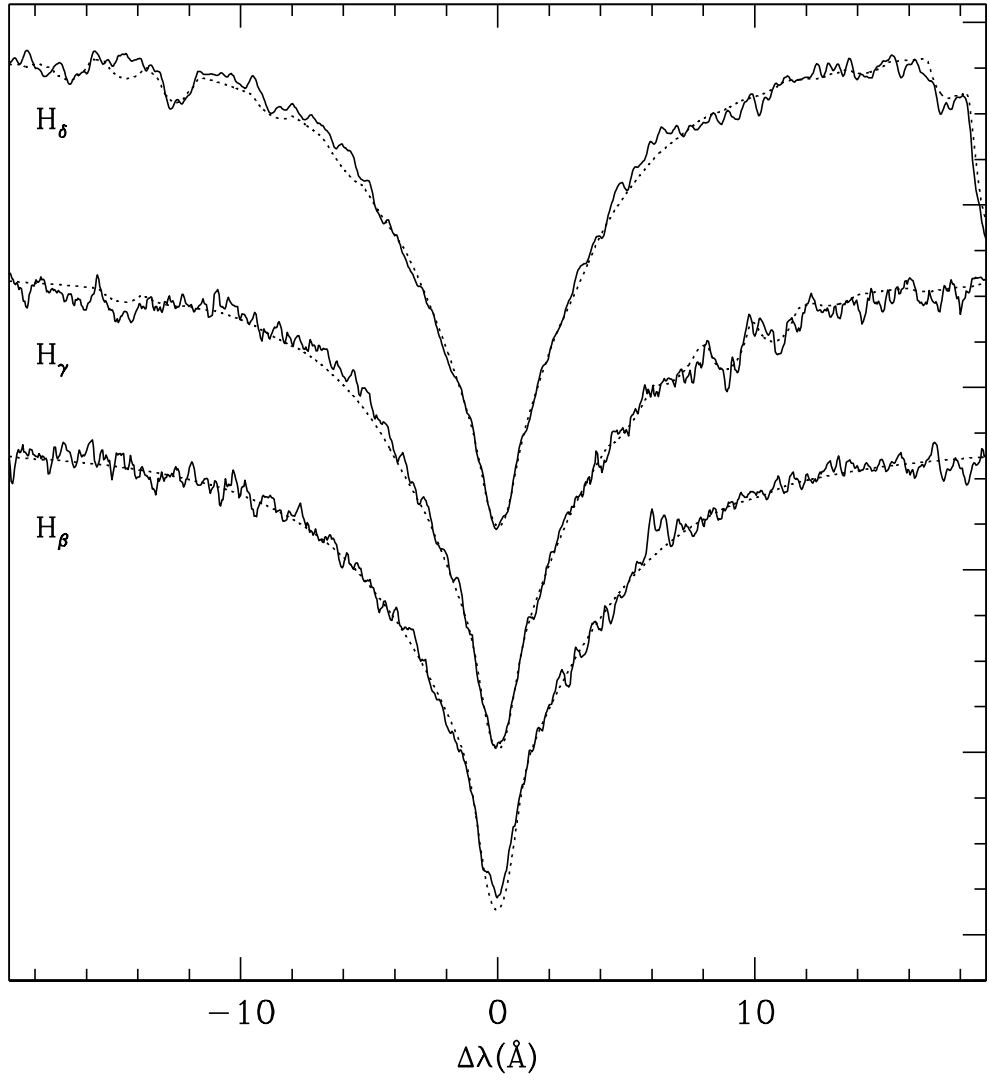}
\caption{Balmer lines of HE 0437--5439. The {\sc tlusty} model spectrum
(dotted line) for the derived parameters is overplotted, illustrating
the good fit to the Balmer wings.}
\label{fig:Balmer}
\end{figure}

We derived the metallicity and atmospheric parameters of HE 0437--5439
using the non-LTE {\sc Tlusty} model atmosphere grid \citep{Hubeny95}
described by \citet{Dufton05}~\footnote{See also
http://star.pst.qub.ac.uk}, which is appropriate for the analysis of
B-type stars. The results for the parameters (effective temperatures,
$T_{\rm eff}$; surface gravity, $g$, in units of cm\,s$^{-2}$;
microturbulence, $\xi$; projected rotational velocity, $v\sin i$) and
chemical abundance of HE 0437--5439 are presented in
Table~\ref{t_abunds}. We estimated the effective temperature from the
ionization balance of \ion{Si}{2} to \ion{Si}{3}, the surface gravity
from fitting the wings of the Balmer lines and the microturbulence from
minimizing the scatter in the abundances of the \ion{Si}{3} triplet of
lines at 4560\AA. The projected rotational velocity of the star was
estimated by fitting rotationally broadened theoretical profiles to 14
strong metal lines and a mean value of 55$\pm$1 km s$^{-1}$ was
obtained. The fits of rotationally broadened profiles to the Balmer
lines are shown in Figure~\ref{fig:Balmer}.

\begin{figure}[hbt]
\epsscale{1.0}
\plotone{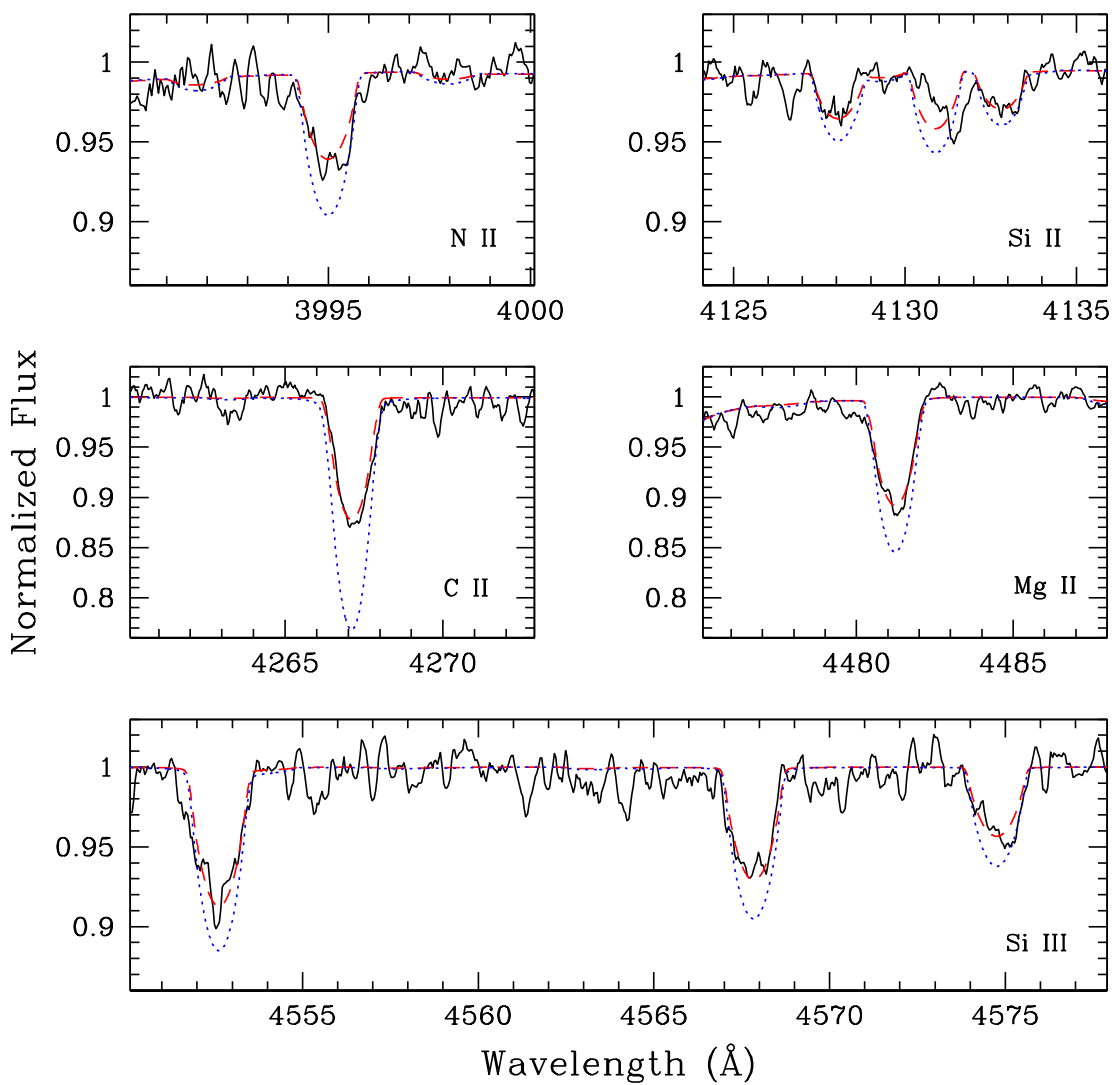}
\caption{Representative metal lines in HE 0437--5439 (solid line) used
to derive abundances. {\sc tlusty} model spectra for half-solar (dashed
line) and solar (dotted line) metallicity are overplotted, illustrating
that the metallicity of HE 0437--5439 is half-solar.}
\label{fig:metals}
\end{figure}

To derive the abundances, we measured the equivalent widths of the metal
absorption lines (C, N, O, Mg and Si). The associated errors were
calculated following the methods of \citet{Hunter07} and include both
random errors (measurement and atomic data uncertainties) and systematic
errors from the adopted atmospheric parameters. We adopted the following
uncertainties: 1,000\,K in effective temperature, 0.2\,dex in surface
gravity and 3\,km\,s$^{-1}$ in microturbulence \citep[for further
details see][]{Hunter07}. Table~\ref{t_abunds} gives in parentheses the
number of lines used for each element. Further, it lists the LMC
baseline abundance (Z=0.5\,\zsun) from \citet{Hunter07} and the baseline
solar abundances from \citet{Asplund05}, for comparison. Our analysis
follows identical methods to \citet{Hunter07}, i.e. we use the same
absorption lines, atomic data and techniques and hence our abundance
neasurements are directly comparable. The carbon abundance from the
4267\AA\ line has been corrected by 0.34\,dex following their
method. The reason for this correction is that the carbon model atom
used in the adopted {\sc Tlusty} model atmosphere grid is relatively
simple compared to that used in more detailed analyses
\citep[e.g.,][]{Sigut96,Nieva06,Nieva07} and leads to a discrepancy
between the carbon abundance derived from the 4267\AA\ line and that
from \ion{H}{2} regions \citep{Hunter07}. The correction is dependent
upon the atmospheric parameters and we therefore note that while our
absolute carbon abundance should be treated with caution, it is
comparable with the LMC stars analysed by \citet{Hunter07}. The adopted
metallicity of the grid (LMC in this case) has a negligible effect on
the derived abundances \citep[$<0.10$ dex, from][]{Hunter07}. The mild
nitrogen enrichement is not uncommon as the enrichment of core processed
material has often been observed in B-type stars \citep[see for
example][]{Walborn70, Dufton72, Gies92, Dufton05, Korn02, Venn99}. In
Figure~\ref{fig:metals}, we plot representative metal lines used to
determine abundances, along with {\sc Tlusty} model spectra at solar and
half-solar metallicities, showing that the half-solar metallicity model
correctly reproduces the metal lines. Clearly, the chemical abundance
derived for HE 0437--5439 indicates an origin from the LMC.

We estimate a slightly higher mass of 9.0 $\pm$ 0.5 $M_{\sun}$ for HE
0437--5439 than \citet[][see their Figure~3]{Edelmann05}, using our new
parameters and the evolutionary models of \citet{Schaerer93}. We also
recalculate the distance to HE 0437--5439 following
\citet{Edelmann05}. With the new values for the magnitude, mass and
surface gravity we obtain a longer distance of $73^{+6}_{-5}$ kpc. A
star of this mass, B2 IV-III spectral type and LMC metallicity should
have an age less than the value estimated by \citet{Edelmann05} for an
$8\, \msun$ main-sequence early-B star at this metallicity, i.e. 35 Myr.
Using the best fit {\sc Tlusty} model spectrum, we confirm the radial
velocity of the star to be 723 $\pm$ 2 km s$^{-1}$ with the IRAF
$rvsao.xcsao$ task. We note that HE 0437--5439 is the only HVS so far
with an accurate rotational velocity measurement: 55 $\pm$ 1 km
s$^{-1}$. \citet{Hansen07} predicted that the rotational velocities of
HVSs ejected by the \citet{Hills88} scenario (originating in binaries)
should be lower than those of single stars of the same spectral
type. Our measurement confirms a low $v\sin i$ for HE 0437--5439.

\section{Discussion} \label{sec:sum}

We measure the chemical abundance of the hypervelocity star HE
0437--5439 and find that it has half-solar metallicity, thus
establishing its origin in the LMC. We can therefore rule out Galactic
origin, because stars in the Galactic center have been found to have
solar or supersolar metallicities \citep[e.g.][]{Carr00, Ramirez00,
Najarro04, Wang06, Cunha07}. A scenario of ejection from the
low-metallicity outskirts of the Galactic disk also suffers from the age
paradox \citep[see Figure~4 in][]{Edelmann05}. Our non-variable
photometry further renders a scenario with a contact binary blue
straggler ejected from our Galaxy unlikely.

HE 0437--5439 must therefore have been ejected from the LMC. A
noteworthy implication of our result is the existence of a massive black
hole in the LMC, as suggested by \citet{Gualandris07}. According to the
simulations these authors performed, an intermediate mass black hole
(IMBH) of mass $\ge1000\,\msun$ is required to eject HE 0437--5439 from
the LMC. A three-body interaction involving a stellar mass black hole,
while still possible, is very unlikely. \citet{Gualandris07} find NGC
2004 and NGC 2100 to be the best candidate hosts of an IMBH in the LMC,
based on criteria of age, density and mass. Further work is necessary to
determine the location of the massive black hole in the LMC.

In summary, by establishing the origin of HE 0437--5439, we find
dynamical evidence for the existence of a hitherto undetected massive
black hole in the LMC.

\acknowledgments{We thank the referee, Rolf Kudritzki, for detailed
comments that significantly improved this manuscript. We particularly
thank K. Stanek for motivating us to monitor this star and for useful
discussions and comments on the manuscript. Furthermore, we thank
J. Hartman for his fluxtomag script, I. Ribas for his script to merge
echelle orders, N. Walborn for confirming our spectral classification,
J. Greene for discussions on intermediate mass black holes, P.L. Dufton
for discussions on the {\sc Tlusty} non-LTE model atmosphere
calculations, and I. Thompson for helpful comments on the manuscript.
AZB acknowledges research and travel support from the Carnegie
Institution of Washington through a Vera Rubin Fellowship. MLM
acknowledges support provided by NASA through Hubble Fellowship grant
HF-01210.01-A awarded by the Space Telescope Science Institute, which is
operated by the Association of Universities for Research in Astronomy,
Inc., for NASA, under contract NAS5-26555. IH acknowledges financial
support from the UK Science and Technology Facilities Council (STFC).}


\clearpage
\begin{deluxetable}{llcc}
\tablewidth{0pc}
\tablecaption{\sc Parameters and Abundances For HE 0437--5439$^{\rm
a}$.}
\tablehead{
\colhead{Parameter} & \colhead{HE 0437--5439} & \colhead{LMC} &
\colhead{Solar}\\ \colhead{} & \colhead{} & \colhead{Abund.} &
\colhead{Abund.}} 
\startdata
T$_{\rm eff}$ (K)        & 21,500$\pm$ 1,000 	    &         &       \\
$\log(g)$ (dex)           & 3.70 $\pm$0.2    	    &         &       \\
$\xi$ (km\,s$^{-1}$)     & 2$\pm$3        	    &         &       \\
$v\sin i$ (km\,s$^{-1}$) & 55$\pm$1      	    &         &       \\
\hline
\ion{C}{2}              &7.79$\pm$0.13 ( 1)& 7.75    & 8.39  \\
\ion{N}{2}              &7.30$\pm$0.24 ( 1)& 6.90    & 7.78  \\
\ion{O}{2}              &8.44$\pm$0.33 (13)& 8.35    & 8.66  \\
\ion{Mg}{2}             &7.10$\pm$0.18 ( 1)& 7.05    & 7.53  \\
\ion{Si}{2}             &7.17$\pm$0.21 ( 2)& 7.20    & 7.51  \\
\ion{Si}{3}             &7.18$\pm$0.34 ( 3)& 7.20    & 7.51  \\

\enddata
\label{t_abunds}
\tablenotetext{a}{See \S\ref{sec:metal} for details.}
\end{deluxetable}

\end{document}